\documentclass{biometrika-ArXiv}
\pdfoutput=1
\usepackage{amsmath}

\usepackage{times}
\usepackage[pdftex]{graphicx}
   \usepackage{epstopdf}
\usepackage{bm,amssymb,amsfonts,url, xcolor,float}
            \usepackage[plain,noend]{algorithm2e}

\makeatletter
\renewcommand{\algocf@captiontext}[2]{#1\algocf@typo. \AlCapFnt{}#2} % text of caption
\def\@algocf@capt@plain{top}
\renewcommand{\algocf@makecaption}[2]{%
  \addtolength{\hsize}{\algomargin}%
  \sbox\@tempboxa{\algocf@captiontext{#1}{#2}}%
  \ifdim\wd\@tempboxa >\hsize%     % if caption is longer than a line
    \hskip .5\algomargin%
    \parbox[t]{\hsize}{\algocf@captiontext{#1}{#2}}% then caption is not centered
  \else%
    \global\@minipagefalse%
    \hbox to\hsize{\box\@tempboxa}% else caption is centered
  \fi%
  \addtolength{\hsize}{-\algomargin}%
}
\makeatother

\def\bco{\iffalse}
\def\cov{{\rm cov}}
\def\var{{\rm var}}
\def\ci{\cite}
\def\cp{\citep}
\DeclareMathOperator*{\argmin}{argmin}

\def\mkF{\mathfrak{F}}
\def\inv{^{-1}}
\def\mc{\mathcal}
\def\mk{\mathfrak}
\def\ra{\rightarrow}

\def\dy{{\rm d}y}
\def\dz{{\rm d}z}
\def\dt{{\rm d}t}
\def\ds{{\rm d}s}
\def\du{{\rm d}u}

\def\i01{\int_0^1}
\def\DT{\mathcal{D}}

\def\I{\mc{I}}
\newcommand{\ed}{\end{document}}

\begin{document}

%\jname{Biometrika}
%% The year, volume, and number are determined on publication
%\jyear{2016}
%\jvol{103}
%\jnum{1}
%% The \doi{...} and \accessdate commands are used by the production team
%\doi{10.1093/biomet/asm023}
%\accessdate{Advance Access publication on 31 July 2016}

%% These dates are usually set by the production team
%\received{April 2012}
%\revised{October 2015}

%% The left and right page headers are defined here:
\markboth{A. Petersen \and H. G. M\"uller}{Wasserstein Covariance}

%% Here are the title, author names and addresses
\title{Wasserstein Covariance for Multiple Random Densities}

\author{ALEXANDER PETERSEN}
\affil{Department of Statistics and Applied Probability, University of California, Santa Barbara, California 93106, U.S.A. \email{petersen@pstat.ucsb.edu}}

\author{\and HANS-GEORG M\"ULLER}
\affil{Department of Statistics, University of California, Davis, California 95616, U.S.A. \email{hgmueller@ucdavis.edu}}

\maketitle

\begin{abstract}
A common  feature of methods for analyzing samples of probability density functions is that they respect the geometry inherent to the space of densities.  Once a metric is specified for this space, the Fr\'echet mean is typically used to quantify and visualize the average density from the sample.  For one-dimensional densities, the Wasserstein metric is popular due to its theoretical appeal and interpretive value as an optimal transport metric, leading to the Wasserstein--Fr\'echet mean or barycenter as the mean density.  We extend the existing methodology for samples of densities in two key directions.  First, motivated by applications in neuroimaging, we consider dependent density data, where a $p$-vector of univariate random densities is observed for each sampling unit.  Second, we introduce a Wasserstein covariance measure and propose intuitively appealing estimators for both fixed and diverging $p$, where the latter corresponds to continuously-indexed densities. We also give theory demonstrating consistency and asymptotic normality, while accounting for errors introduced in the unavoidable preparatory density estimation step.  The utility of the Wasserstein covariance matrix is demonstrated through  applications to  functional connectivity in the brain using functional magnetic resonance imaging data and to the secular evolution of mortality for various countries. 
\end{abstract}


\section{Introduction}

\label{sec: intro}

The analysis of samples of density functions or distributions is  
an important and challenging problem for modern statistical practice \cp{deli:11}. Examples include distributions of age at death or mortality for different countries, warping functions, or  distributions of voxel-to-voxel correlations of functional magnetic resonance imaging signals for a sample of subjects.  While functional  principal component analysis using cross-sectional averaging can be directly applied for samples of density functions \cp{knei:01},  more recently techniques have been developed	 that incorporate the geometric constraints inherent to the space of density functions. A popular metric  
for data where each data atom corresponds to a randomly sampled distribution or density is the Wasserstein metric, both for its theoretical appeal and its convincing  empirical performance in various applications \cp{bols:03, mull:11:4, bigo:16,bigo:17,pana:16}.

We consider samples of density data where multiple random densities per subject are observed, that is, repeated realizations of a stochastic process defined on $\mathcal{D}^p$, where $\mathcal{D}$ is a space of one-dimensional density functions. For  $p = 1$, a variety of methods have been proposed, focusing on concepts of mean, modes of variation and dimensionality reduction \cp{deli:11, mull:16:1, mena:16:2}.  However, generalizations of these methods for $p>1$, where one has a vector of densities,  have not yet been developed, even though such data arise in various applications.  An essential first step is the extension of the mean and variance concepts for samples of densities to a vector of means and covariance matrix. 
We demonstrate here the usefulness of these concepts for applications. The type of data we consider is different from data where one observes, for each subject, a random sample of identically distributed $p$-vectors, for which the joint distribution is of interest.  Instead, we focus on the joint modeling of $p$ univariate densities.  

Using the Wasserstein geometry of optimal transport on the space of densities, we extend the concepts of Fr\'echet mean and variance to a measure of Wasserstein covariance  
between component densities for a $p$-variate density-valued process. As one does not observe the densities themselves but rather samples of univariate data that they generate, a preliminary density estimation step is necessary and is taken into account in our analysis.  Our theoretical arguments show that the population Wasserstein covariance can be estimated consistently, with a limiting Gaussian distribution under sufficient conditions for the estimation error associated with the preliminary estimation step to be negligible.  Motivated by one of the applications, we also consider continuously varying densities, where one observes a discretized version of a continuously evolving density-valued process,
 similar to densely observed repeated functional data \cp{park:15,mull:17:4}.  For this situation we target a Wasserstein covariance surface and develop theory for its consistent estimation while still accounting for errors in density estimation.

The utility of the proposed methodology  is demonstrated through the analysis of functional magnetic resonance imaging and mortality data.  In the application to brain imaging, we investigate differences in intra-regional functional connectivity between Alzheimer's patients, cognitively normal subjects, and a third group of individuals  diagnosed with mild cognitive impairment, a sign of increased risk for developing dementia.  Here, intra-regional  functional connectivity is quantified by the distribution of voxel-to-voxel correlations within a neighborhood, called a functional hub, corresponding to a small region in the brain. Multiple densities per subject are obtained for a number of hubs simultaneously, leading to a random vector of densities obtained for each subject in the sample. In the mortality application,  we consider distributions of age at death over a range of calendar years, and compare the resulting Wasserstein covariance surfaces between a group of Eastern European countries and a second group of other nations.    

\section{Wasserstein Covariance}
\label{sec: random_densities}

\subsection{The Wasserstein Metric and Geometry}
\label{ss: wass_metric}

Let $\mc{D}$ be a class of one-dimensional densities such that $\int_{\mathbb{R}}u^2 f(u)\du < \infty$ for all $f \in \mc{D}.$ For $f,\, g \in \mc{D}$, suppose $Y\sim f$ and consider the collection of nondecreasing maps \mbox{$T^*:\mathbb{R}\rightarrow\mathbb{R}$,} known as transports, such that $T^*(Y) \sim g.$  The optimal transport problem that leads to the Wasserstein metric is to find the transport that minimizes 
\begin{equation}
\label{eq: optimal_transport}
\int_{\mathbb{R}} \left\{T^*(u) - u\right\}^2 \, f(u)\du,
\end{equation}
and the solution is known to be $T = G\inv \circ F$, where $F$ and $G$ are the distribution functions of $f$ and $g$, respectively \cp{ambr:08}.  The resulting squared Wasserstein distance is 
\begin{equation}
\label{eq: wass_metric}
d_W^2(f, g) = \int_{\mathbb{R}}\left\{T(u) - u\right\}^2f(u)\du. 
\end{equation}
This metric  arises from a local inner product \cp{ambr:08}.  If  $g_j \in \mc{D}$ ($j = 1,2)$ have distribution functions $G_j,$ the optimal transports $T_j = G_j\inv \circ F$ for given $F$ 
reside in the tangent space of $f$, and for each $f$ an  inner product between $T_1$ and $T_2$ can be defined by 
\begin{equation}
\label{eq: wass_ip}
\langle T_1, T_2 \rangle_{f} = \int_{\mathbb{R}} \left\{T_1(u) - u\right\}\left\{T_2(u) - u\right\}\, f(u)\du,
\end{equation}
so that $d_W(f, g_j)^2 = \langle T_j, T_j\rangle_{f} = \langle T_j\inv, T_j\inv \rangle_{g_j}.$

\subsection{Wasserstein Mean, Variance and Covariance}
\label{ss: wf_mean}

For a random density $\mk{F}$ in $\DT$, 
its Fr\'echet mean and Fr\'echet variance \cp{frec:48} are natural tools for relating the distributional properties of $\mk{F}$ to the Wasserstein geometry of $\DT$ in terms of first and second order behaviour.  When the space $\DT$ is endowed with the metric $d_W$, the Wasserstein--Fr\'echet, or simply Wasserstein, mean and variance of $\mk{F}$ are
\begin{equation}
\label{eq: wass_mean_var}
f_\oplus = \argmin\limits_{f \in \DT} E\left\{d_W(\mathfrak{F}, f)^2\right\}, \quad \var_\oplus(\mk{F}) = E\left\{d_W(\mathfrak{F}, f_\oplus)^2\right\}.
\end{equation}
For a single density process $\mk{F},$ the theoretical and practical properties of the Wasserstein mean have been  thoroughly investigated 
\cp{bols:03,mull:11:4,pana:16,bigo:17}, and recently the Wasserstein variance was  adopted to quantify variability explained when performing dimension reduction for densities \cp{mull:16:1}.  To quantify the dependence between two random densities, we propose here the extension of these concepts to a Wasserstein covariance measure. 

For a bivariate density process $(\mathfrak{F}_1,\mathfrak{F}_2)$, where the  $\mathfrak{F}_j$ are random elements of $\mc{D}$, $j=1,2,$ denote by $F_{\oplus,j}$ the distribution function of the Wasserstein mean of $\mathfrak{F}_j$  \eqref{eq: wass_mean_var} and the corresponding density by  $f_{\oplus,j}$. 
The random optimal transport from $f_{\oplus,j}$ to $\mathfrak{F}_j$ is $T_j = F_j\inv \circ F_{\oplus ,j},$ where $F_j$ is the distribution function of $\mk{F}_j,$ so that the Wasserstein variances are
$$
\var_\oplus(\mk{F}_j) =  E\left\{d_W(\mk{F}_j, f_{\oplus,j})^2\right\} = E\left[\int_{\mathbb{R}}\left\{T_j(u) - u\right\}^2f_{\oplus,j}(u)\du\right] = E\left(\langle T_j,T_j\rangle_{f_{\oplus,j}}\right).
$$
This  suggests defining a Wasserstein covariance measure as an expected inner product between $T_1$ and $T_2.$  Since in general $f_{\oplus,1} \neq f_{\oplus,2},$ these transports reside in different tangent spaces. Adopting a common technique for manifold-valued data 
 \cp{yuan:12},  we push  $T_1$ to a new transport map $\tilde{T}_1$ in the tangent space of $f_{\oplus,2}$ by a parallel transport map.  Intuitively, as $T_1$ is transported to $\tilde{T}_1$  along the geodesic connecting $f_{\oplus,1}$ to $f_{\oplus,2},$ its angle with the geodesic is preserved.  In the Wasserstein geometry,  setting $T_{\oplus,12} = F_{\oplus,1}\inv\circ F_{\oplus,2}$, one obtains  $\tilde{T}_1 = T_1\circ T_{\oplus,12} - T_{\oplus,12} + \textrm{id},$ where $\textrm{id}$ is the identity map.  Due to symmetry of this operation, if $\widetilde{T}_2 = T_2\circ T_{\oplus,12}\inv - T_{\oplus,12}\inv + \textrm{id},$ then $\langle \widetilde{T}_1, T_2 \rangle_{f_{\oplus,2}} = \langle T_1, \widetilde{T}_2\rangle_{f_{\oplus,1}}.$  Thus, the Wasserstein geometry motivates
\begin{equation}
\label{eq: wass_cov}
\cov_\oplus(\mk{F}_1, \mk{F}_2) = E\left(\langle \widetilde{T}_1, T_2 \rangle_{f_{\oplus,2}}\right) = E\left(\langle T_1, \widetilde{T}_2 \rangle_{f_{\oplus,1}}\right) = \cov_\oplus(\mk{F}_2, \mk{F}_1)
\end{equation}
as the Wasserstein covariance between the two random densities.

\subsection{Expression in Terms of Quantile Functions}
\label{ss: wqnt}

The well known fact that for one-dimensional densities, the Wasserstein geometry is closely related to quantile functions \cp{vill:03,pana:16,mull:16:1} leads to  a second characterization of Wasserstein covariance, which is useful for practice. The change of variable $t = F(u)$ applied to \eqref{eq: wass_ip} gives the alternative expression
\begin{equation}
\label{eq: wass_quant}
\langle T_1, T_2 \rangle_f = \int_0^1\left\{F\inv(t) - G_1\inv(t)\right\}\left\{F\inv(t) - G_2\inv(t)\right\}\dt,
\end{equation}
so that $d_W^2(f, g) = \i01 \{F\inv(t) - G\inv(t)\}^2\dt.$ We can then express the quantities in \eqref{eq: wass_mean_var} and \eqref{eq: wass_cov} in terms of the random quantile functions $F_j\inv.$  The Wasserstein means $f_{\oplus,j}$ are characterized by their quantile functions $F_{\oplus,j}\inv(t) = E\{F_j\inv(t)\}$ ($0 \leq t \leq 1;\, j = 1,2),$ and the Wasserstein variances and covariances are
\begin{equation} \label{eq:c}
\cov_{\oplus}(\mk{F}_j,\mk{F}_k) = E\left[\i01 \left\{F_j\inv(t) - F_{\oplus,j}\inv(t)\right\}\left\{F_k\inv(t) - F_{\oplus,k}\inv(t)\right\}\dt\right] \,\, (j,k = 1,2).
\end{equation}

Expressions \eqref{eq: wass_quant}, \eqref{eq:c}   reveal a connection to functional data analysis.  Viewing $(F_1\inv,F_2\inv)$ as bivariate functional data, key objects are the mean functions $F_{\oplus,j}\inv(t)$ and covariance surfaces $$\mc{C}_{jk}(s, t) = \cov\left\{F_j\inv(s),F_k\inv(t)\right\} \,\,(j,k = 1,2;\,  0 \leq s , t \leq 1),$$ that characterize the first- and second-order behavior of the processes \cp{li:10}.  Writing $\mc{M}_{jk}$ for  the integral operator with kernel $\mc{C}_{jk},$ Fubini's theorem implies that $$\cov_{\oplus}(\mk{F}_j,\mk{F}_k) = \i01 \mc{C}_{jk}(t, t)\dt = \textrm{Tr}(\mc{M}_{jk}),$$ where $\textrm{Tr}(\cdot)$ is the operator trace. Accordingly,  the Wasserstein variance of each component distribution can be interpreted as a summary of the  variability in the quantile process, and Wasserstein covariance as a summary of  covariability.

Using quantile functions has two major advantages when multiple densities are observed per subject.  First, the derived notions of mean, variance, and covariance have geometric interpretations in the manifold induced by the Wasserstein metric in the 
space of distributions, as in \eqref{eq: wass_mean_var} and \eqref{eq: wass_cov}.  Second, quantile functions  always have the same support $[0,1]$ regardless of the distributional supports of the $\mk{F}_j,$ so that the Wasserstein covariance remains well-defined even when the latter differ.  In contrast, attempts to define similar covariance summaries based on cross-covariance operators of density or compositional representations are bound to fail when distributional supports differ.  For example, if $\mc{G}_{jk}$ is the ordinary cross-covariance operator between densities $\mk{F}_j$ and $\mk{F}_k$, the operator trace is well-defined only when the supports coincide.  Even when they are the same, taking $\textrm{Tr}(\mc{G}_{jk})$ as a summary covariance measure has no intuitive geometric meaning. 

\subsection{Wasserstein Covariance Matrices and Kernels}
\label{ss: wcov_ext}

Considering  a $p$-variate density process $\mathfrak{F} = (\mk{F}_1,\ldots,\mk{F}_p)$ with component Wasserstein means and variances $f_{\oplus, j}$ and $\var_\oplus(\mk{F}_j)$, $j = 1,\ldots,p,$ 
we initially  assume that  $p$ is fixed,  as in the brain imaging example in Section~\ref{ss: brain}.   
The Wasserstein covariance matrix for $\mk{F}$ is the $p\times p$ matrix with elements
\begin{equation}
\label{eq: wcov_mat}
\left(\Sigma_\oplus\right)_{jk} = \cov_\oplus(\mk{F}_j,\mk{F}_k), 
\end{equation}
which is easily seen to be a valid covariance matrix. 

Motivated by the mortality example in Section~\ref{ss: mortality}, suppose the components of $\mk{F}$ are indexed by a continuous variable $0 \leq y_j \leq 1$, which might represent time. 
To model the setting of repeatedly observed densities that are measured densely in time, we allow $p \rightarrow \infty$. Then $\mk{F}$ can be thought of as a discretized version of an unobservable process $\mc{F}(y),$ $0 \leq y \leq 1,$ with $\mk{F}_j = \mc{F}(y_j).$   
We target the Wasserstein mean surface $f_\oplus(\cdot; y)$ and covariance kernel
\begin{equation}
\label{eq: wcov_kernel}
\Sigma_{\oplus}(y,z) = \cov_{\oplus}\left\{\mc{F}(y), \mc{F}(z)\right\} \quad (0 \leq y,z \leq 1).
\end{equation}

\bco{Denoting the quantile process for $\mk{F}_j$  by  $X_j$, 
the distribution function of  $f_{\oplus, j}$ by $F_{\oplus,j}$   and the random optimal transport map for $\mk{F}_j$ by $T_j = X_j \circ F_{\oplus, j}$, we have  $E\left\{X_j(t)\right\} = F_{\oplus, j}\inv(t)$ and $E\left\{T_j(u)\right\} = u$.  With
\begin{equation}
\label{eq: cross_cov}
\mc{C}_{jk}(s, t) = \cov\left\{X_j(s), X_k(t) \right\}, \quad \mc{G}_{jk}(u, v) = \cov\left\{T_j(u), T_k(v)\right\},
\end{equation}
a canonical generalization of \eqref{eq: wass_var} to the Wasserstein covariance between components $j$ and $k$ is the total \emph{covariability} between quantile processes
\begin{equation}
\label{eq: wcov}
\left(\Sigma_\oplus\right)_{jk} = \i01\mc{C}_{jk}(t, t)\dt.
\end{equation}
Lastly, we can reformulate this covariance measure in terms of the crosscovariance $\mc{G}_{jk}$ by utilizing the optimal transport between mean measures, $T_{\oplus, jk} = F_{\oplus,j}\inv \circ F_{\oplus, k}$.  Using the change of variables $u = F_{\oplus, j}\inv(t)$ and $v = F_{\oplus, k}\inv(t)$, with $\tilde{u}(u) = T_{\oplus,kj}(u)$ and $\tilde{v}(v) = T_{\oplus,jk}(v),$}\fi

\section{Estimation of Wasserstein Covariance Objects}
\label{sec: est_theory}

\subsection{Density Estimation}
\label{ss: dens_est}

While we defined the Wasserstein covariance for fully observed densities, in reality these densities are rarely if ever directly observed. Rather, the data actually available are collections of scalar random variables $W_{ijr}$ ($i = 1,\ldots,n$;  $j = 1, \ldots p$; $r= 1,\ldots,N_{ij}$), where $n$ is the number of subjects $i$, $p$ is the number of densities or distributions $j$ per subject, and $N_{ij}$ is the number of independent observations $r$ distributed according to the $j$-th density that are available for the $i$-th subject and may vary across $i,j$.  The observed data can be viewed as resulting from  two independent random mechanisms, where the first random mechanism generates the independent vectors of densities $f_i = \{f_{i1},\ldots,f_{ip}\}$ $(i = 1,\ldots,n$), while the second generates the observations that are  sampled from these distributions, $W_{ijr} \sim f_{ij}$. The $W_{ijr}$ are all  independent and for each fixed $(i,j)$ are also identically distributed. 

Obtaining density estimates $\hat{f}_{ij}$ from the observed data   $W_{ijr} $ and using these as proxies for the $f_{ij}$, for the asymptotic analysis we need to 
address the challenge that these estimates are noisy and deviate from the true density targets. Since the targets $\Sigma_\oplus$ can be expressed as integrated moments of the multivariate quantile process, an obvious route would be to estimate the empirical quantile functions and proceed by averaging.  However, this has some practical drawbacks and a 
preferred approach is  to first construct a sample of smooth density estimates  
$\hat{f}_{ij}$, then obtaining smooth distribution functions by integration,  quantile functions as inverse functions and the target quantities from the estimated quantile functions. 

The theoretical analysis of the Wasserstein covariance estimates in Section~\ref{ss: wcov_est} below requires the following assumption, where  a preliminary density estimator is  generically denoted by $\hat{f}$. 
\begin{assumption}
\label{asm: D1}
There is a compact interval $I$ such that, for any $f \in \mc{D},$ its support $D_f$ is a compact interval contained in $I.$  If $W_1,\ldots,W_N$ is a random sample from $f$, the density estimate $\hat{f}$ based on this sample is a probability density function on $D_f$ such that, for some decreasing sequence $b_N = o(1)$ as $N \ra \infty,$
\[
\sup_{f \in \mc{D}} E\left\{d_W(f, \hat{f})\right\} = O(b_N).
\]
\end{assumption}
\noindent 
A density estimator that satisfies Assumption~\ref{asm: D1} is described in \ci{mull:16:1}.  If the support $D_f$ is known and the condition $$\sup_{f \in \mc{D}}\,\sup_{u \in D_f} \max\left\{f(u), 1/f(u), |f'(u)|\right\} < \infty$$ is satisfied, then one may take $b_N = N^{-1/3};$ see Proposition 1 in the Supplementary Material. 
With no assumed uniform lower bound on the random densities, \ci{pana:16} proposed a density estimator for which $b_N = N^{-1/4}.$  

Lastly, because $np$ densities need to be simultaneously estimated, each from data of varying sample sizes  $N_{ij}$, these need to be tied to the number of independent subjects $n$, as follows. 
\begin{assumption}
\label{asm: D2} There exists a sequence $N = N(n)$ such that $\min_{i,j} N_{ij} \geq N$ and $N  \ra \infty$ as $n \ra \infty$.
\end{assumption}
Here $N$ is a uniform lower bound on the sample sizes for the $np$ densities to be estimated that  must diverge with the number of subjects $n$ to ensure uniform consistency of the densities.

\subsection{Wasserstein Covariance Estimation}
\label{ss: wcov_est}

For fixed $p$,  given densities  $f_i = (f_{i1},\ldots,f_{ip})$,  $i = 1,\ldots,n,$ that are independently and identically distributed according to $\mathfrak{F} = (\mkF_1,\ldots,\mkF_p)$, our main goal is to estimate the Wasserstein covariance matrix $\Sigma_\oplus$, with elements defined in (\ref{eq: wcov_mat}). We compute density estimates $\hat{f}_{ij}$, which are then mapped to their quantile function estimates $\hat{X}_{ij} = \hat{F}_{ij}\inv$.
\bco{Starting with  pairs $(u_a, \hat{f}_{ij}(u_a))$,  we first obtain cumulative distribution functions by numerical integration, leading to  $(u_a, \hat{F}_{ij}(u_a))$, where the $u_a$ are gridpoints on $\I_j$.  With $t_a = \hat{F}_{ij}(u_a)$, we then compute discretized quantile function estimates $\hat{F}_{ij}\inv(t_a) = u_a$, which can then be interpolated to a desired, say equispaced, input grid $\tilde{t}_a$ on $[0,1]$. }\fi Write $\hat{X}_{ij}^c = \hat{X}_{ij} - n\inv \sum_{i = 1}^n \hat{X}_{ij}$ and $\hat{\mc{C}}_{jk}(s, t) = n\inv \sum_{i = 1}^n \hat{X}_{ij}^c(s)\hat{X}_{ik}^c(t).$   To target the Wasserstein covariance $\Sigma_{\oplus},$ the results of Section~\ref{ss: wqnt} suggest the estimator
\begin{equation}
\label{eq: wcov_est}
\left(\hat{\Sigma}_\oplus\right)_{jk} = \int_0^1\hat{\mc{C}}_{jk}(t, t)\dt.
\end{equation}
Theorem 1 demonstrates the overall rate of convergence of the Wasserstein covariance estimator, establishing  asymptotic normality when $N$ diverges sufficiently fast. While our focus is on the Wasserstein covariance, the same rate of convergence is also obtained for the full covariance surface $\hat{\mc{C}}_{jk}(s,t)$ as an estimator of $\mc{C}_{jk}(s,t)$; see Theorem 3 in the Supplementary Material, where also auxiliary results and proofs can be found.
\begin{theorem}
\label{thm: wcov_est}
Suppose Assumptions~\ref{asm: D1} and \ref{asm: D2} hold, and that $\mk{F}_j \in \mc{D}$ almost surely ($j = 1,\ldots,p).$ Then $\lVert \Sigma_\oplus - \hat{\Sigma}_\oplus\rVert_F = O_p(n^{-1/2} + b_N),$ where $\lVert \cdot \rVert_F$ denotes the Frobenius norm.  Additionally, if $n^{1/2}b_N$ converges to zero, then there exists a zero-mean $p\times p$ Gaussian matrix $\mathfrak{C}$ such that $n^{1/2}\left(\hat{\Sigma}_\oplus - \Sigma_\oplus\right)$ converges weakly to $\mathfrak{C}.$

\end{theorem}
The covariance structure of $\mathfrak{C}$ is a four-dimensional array defined in the Supplementary Material.  Under regularity conditions, the density estimator in \ci{mull:16:1} satisfies $b_N = N^{-1/3}$ so that asymptotic Gaussianity is obtained for $N = n^q$, $q > 3/2.$ Faster rates of convergence $b_N=N^{-\rho}$, with $1/3 < \rho < 1/2$, can be obtained under additional smoothness conditions and for suitable density estimators, and then weaker conditions on $q$ will suffice. 

In the continuously indexed case, the vectors of densites \mbox{$f_i = (f_{i1},\ldots,f_{ip})$} correspond to discretized versions of independent and identically distributed realizations $\tilde{f}_i(\cdot\,; y)$ of a latent dynamic density surface $\mc{F}(y),$ \mbox{$0 \leq y \leq 1,$} where $f_{ij}$ are independently distributed for different $i$ as $\mc{F}(y_{ij}).$  For simplicity, we require 
\begin{assumption}
\label{asm: obs_times}
The number of observation times $p=p(n)$ satisfies $np\inv = O(1)$, and these are equidistant and common for all subjects, i.e., 
$ y_{ij} = (j - 1)/(p-1), \, 1 \leq j \leq p.$
\end{assumption}
We then estimate $\Sigma_\oplus(y_j,y_k)$ 
by the sample Wasserstein covariance kernel estimator
\begin{equation}
\label{eq: wcov_kernel_est}
\hat{\Sigma}_\oplus(y_j, y_k) = \int_0^1 \hat{C}_{jk}(t, t) \dt,
\end{equation}
followed by linearly interpolating these discretized estimates to obtain $\hat{\Sigma}_{\oplus}(y,z)$ for any \mbox{$0 \leq y,z \leq 1.$}  For  this interpolation to be negligible, we require two additional assumptions.
\begin{assumption}
\label{asm: Sig_L}
There exists a constant $L_1 > 0$ such that
$$
|\Sigma_\oplus(y_1,z_1) - \Sigma_\oplus(y_2,z_2)| \leq L_1(|y_1 - y_2| + |z_1 - z_2|) \quad (0 \leq y_1,y_2,z_1,z_2 \leq 1).
$$
\end{assumption}
\begin{assumption}
\label{asm: G_L}
With $X_i(t\,;y)$ denoting  the random quantile function corresponding to $\tilde{f}_i(\cdot\,; y)$ and $X_i^c(t\,; y) = X_i(t\,; y) - E\{X_i(t\,;y)\},$  
$$
\int_{[0,1]^4} \cov\left\{ X_1^c(s\,;y) X_1^c(s\,; z),  X_1^c(t\,;y) X_1^c(t\,; z)\right\}\ds\,\dt\,\dy\,\dz < \infty.
$$
\end{assumption}

\begin{theorem}
\label{thm: wcov_kernel}
Suppose Assumptions~\ref{asm: D1}--\ref{asm: G_L} hold and that $\mc{F}(y) \in \mc{D}$ almost surely. \bco{Then there is a zero-mean Gaussian process $\mathfrak{S}$ on $L^2[0,1]^2$ such that
$$
\sqrt{n}\left(\tilde{\Sigma}_\oplus - \Sigma_\oplus\right) \rightsquigarrow \mathfrak{S},
$$ 
and
$$
\int_0^1\int_0^1 \left\{\tilde{\Sigma}_\oplus(y, z) - \hat{\Sigma}_\oplus(y, z)\right\}^2 \ds\,\dt = O_p(b_m^2),
$$}\fi
Then $$\int_0^1\int_0^1 \left\{\Sigma_\oplus(y, z) - \hat{\Sigma}_\oplus(y, z)\right\}^2 \dy\,\dz = O_p(n\inv + b_N^2).$$
\end{theorem}

\section{Applications}
\label{sec: applications}

\subsection{Functional Connectivity in Brain Imaging}
\label{ss: brain}

The study of functional connectivity in the brain involves the identification of voxels or regions which exhibit similar behaviour, as quantified by  neuroimaging techniques such as electroencephalography and functional magnetic resonance imaging.  Of special interest are  connections that are present  when  subjects are in the resting state \cp{alle:14}.  Studies of connections between spatially remote regions revealed the so-called default mode network in the resting brain, including nodes of high centrality that have been characterized as functional connectivity hubs \cp{buck:09}.  The strength of connections between neighboring voxels, as opposed to those between remote regions, contains important information related to various biological factors \cp{gao:16} and neurological diseases \cp{zale:12}. The strength of local connections in a particular brain region has been quantified in various ways \cp{toma:10, zang:04}.

\ci{mull:16:1} demonstrated the utility of a relatively simple approach to quantifying  local connectivity within functional connectivity hubs that uses probability density functions formed by smoothing histograms of pairwise temporal correlations between the signals of each voxel within a hub and the signal at its central seed voxel. We implement 
this approach 
for the $p = 10$ hubs in Table 3 of \ci{buck:09}. The resulting 10-dimensional vectors of densities that are obtained for each subject are then analyzed with  the proposed Wasserstein covariance. The  ten hubs of interest are located in the  
left and right parietal lobes, medial superior frontal lobe, medial prefrontal lobe, left and right middle frontal lobes, posterior cingulate/precuneus region, right supramarginal lobe and the left and right middle temporal lobes.  

\begin{figure}[t]
\centering
\includegraphics[scale = 0.6]{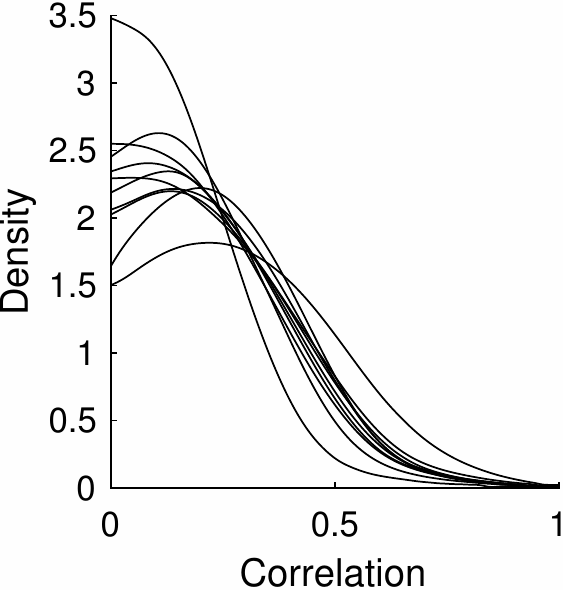} 
\includegraphics[scale = 0.6]{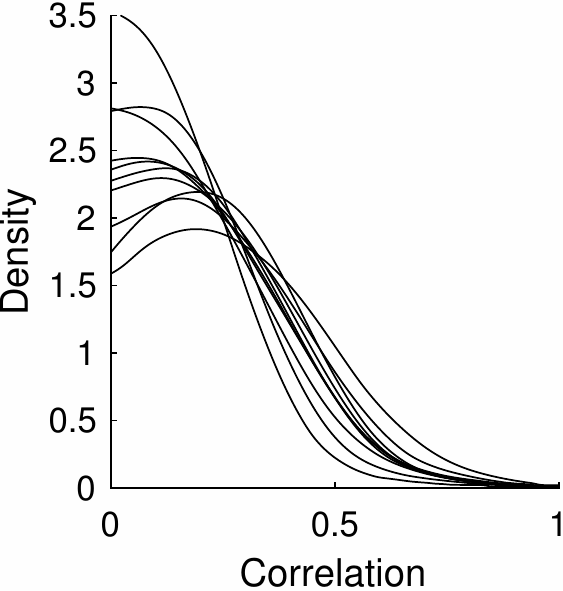}
\includegraphics[scale = 0.6]{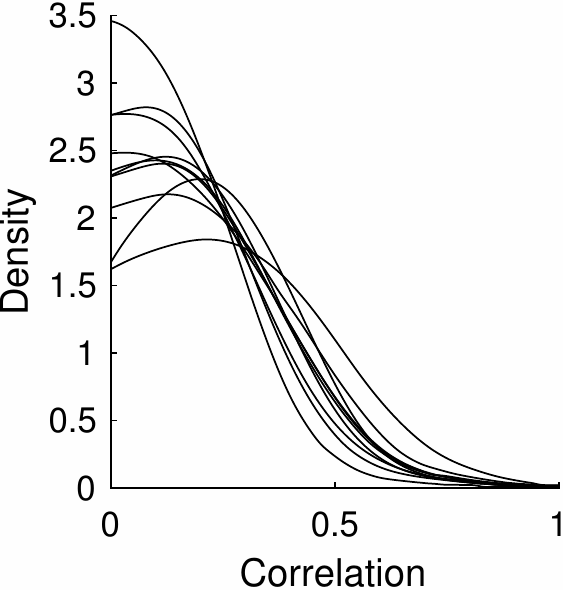}
\caption{Wasserstein means for ten functional connectivity hubs for normal (left), mild cognitive impairment (middle) and Alzheimer's (right) subjects, displayed as densities.  \label{fig: brain_wmean}}
\end{figure}

\begin{figure}[H]
\centering
\includegraphics[scale = 0.7]{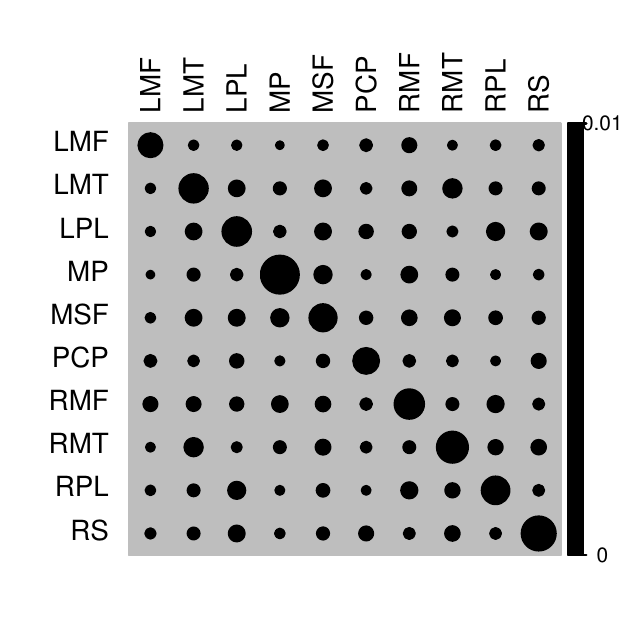}
\includegraphics[scale = 0.7]{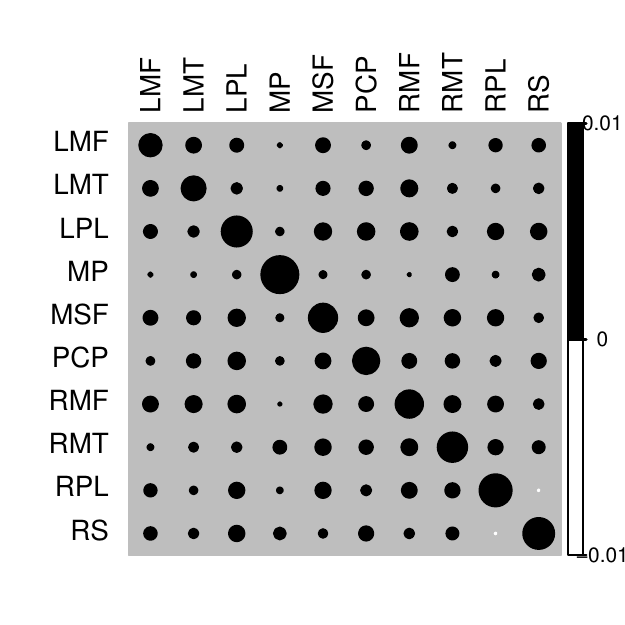}
\includegraphics[scale = 0.7]{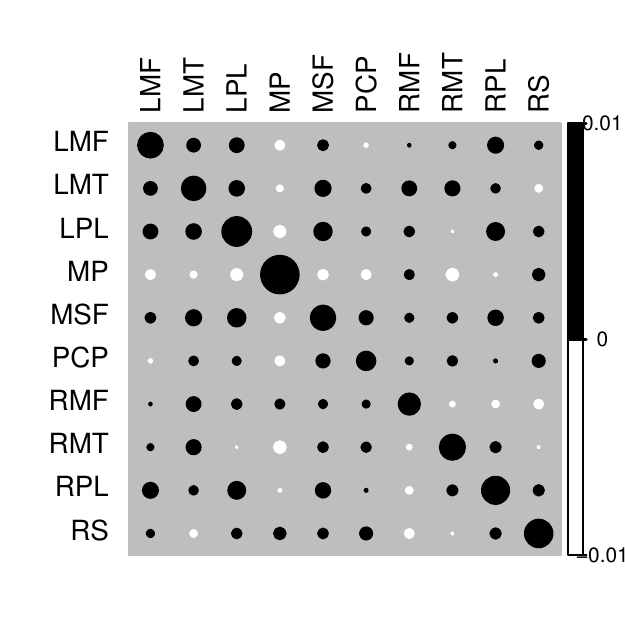} \\
\includegraphics[scale = 0.7]{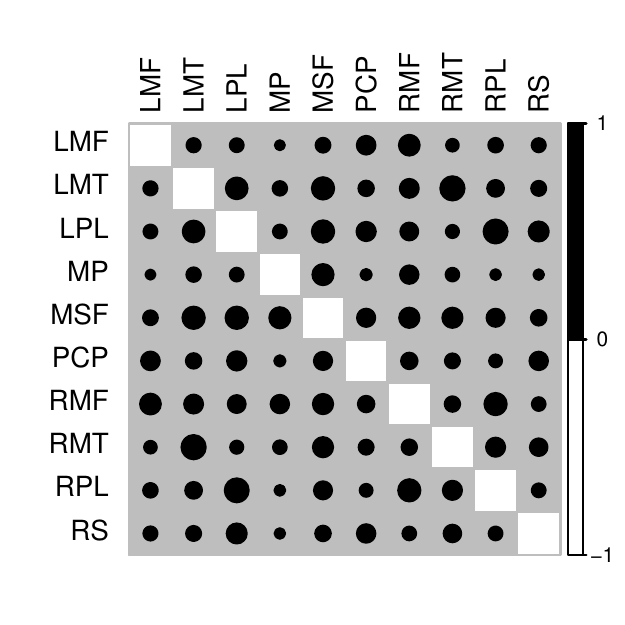} 
\includegraphics[scale = 0.7]{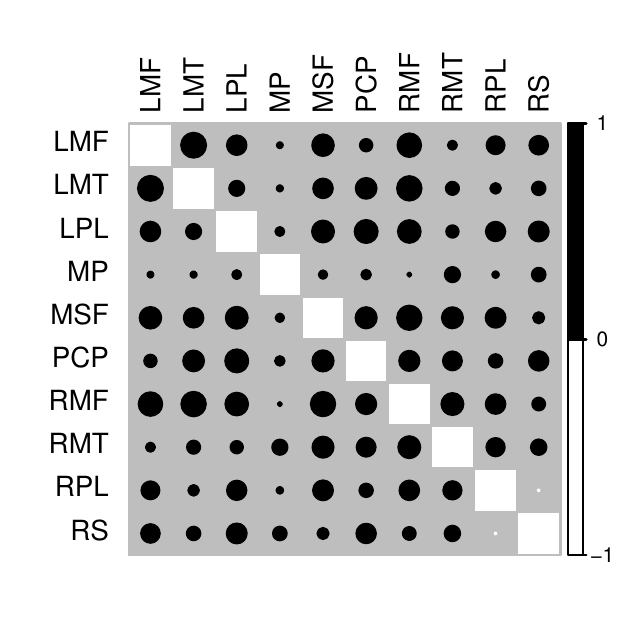}
\includegraphics[scale = 0.7]{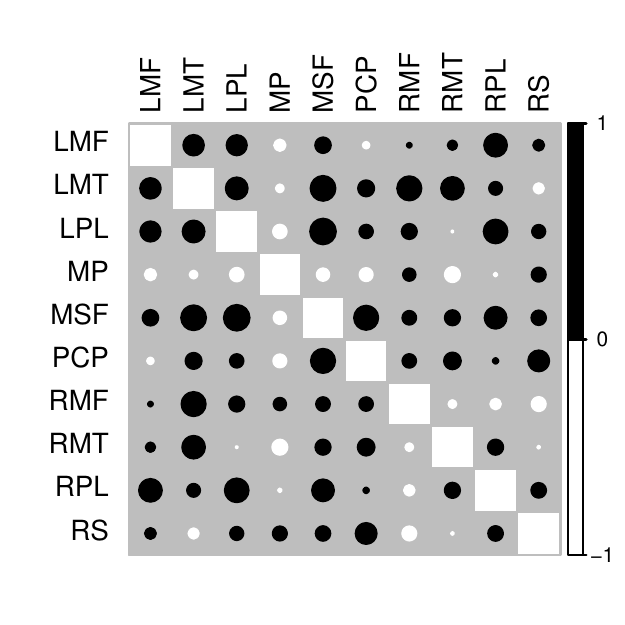}
\caption{Estimated Wasserstein covariance (top row) and correlation (bottom row) matrices for normal (left), mild cognitive impairment (middle) and Alzheimer's (right) subjects. Labels: LMF and RMF (left and right middle frontal), LPL and RPL (left and right parietal), LMT and RMT (left and right middle temporal), MSF (medial superior frontal), MP (medial prefrontal), PCP (posterior cingulate/precuneus) and RS (right supramarginal). Positive (negative) values are drawn in black (white) and larger circles correspond to larger absolute values.\label{fig: wass_corrs}}
\end{figure}

The densities were constructed for  $171$ cognitively normal subjects, $65$ subjects with Alzheimer's disease, and a third group of $86$ subjects with mild cognitive impairment, with details about data preprocessing as in  
\ci{mull:16:12}.  The densities were then used to  compute estimated Wasserstein mean densities $\hat{f}_{\oplus, j}$ for each group separately, which are displayed for each of the 10 hubs in Figure~\ref{fig: brain_wmean}. The various groups show remarkable similarities in terms of the average Wasserstein connectivity distributions across these ten functional hubs.  
In contrast, the Wasserstein covariance  and correlation matrices  in Figure~\ref{fig: wass_corrs} exhibit  clear differences between the  groups and provide valuable additional information.  Correlations are overall stronger for the normal subjects and those with mild cognitive impairment and exhibit different patterns of dependency across  hubs. Complete second order behaviour can be studied through the estimated quantile covariance function estimates $\hat{\mc{C}}_{jk}$, where for $p = 10,$ there are 45 such functions for each group.  We examine the off-diagonal elements for  $j < k$ by plotting slices corresponding to the three quartiles $s = 0.25, 0.5, 75$; see the Supplementary Material. 

We can also test for group differences in Wasserstein covariance using bootstrap samples obtained by centering all quantile functions with respect to their group means. To construct a bootstrap sample under the null hypothesis that all groups have the same Wasserstein covariance matrix, we simply center each multivariate quantile process at its corresponding group Wasserstein mean. 
By pooling these centered processes, a bootstrap sample is obtained by sampling with replacement and then dividing into three groups of proper size. With estimates $\hat{\Sigma}_\oplus^n$, $\hat{\Sigma}_\oplus^m$ and $\hat{\Sigma}_\oplus^a$ for the groups of normal, mildly  cognitively  impaired, and Alzheimer's subjects, the test statistic for the global null hypothesis that all groups are equal is  $$S = \left\lVert \log\hat{\Sigma}_\oplus^n - \log\hat{\Sigma}_\oplus^m\right\rVert_F^2 + \left\lVert \log\hat{\Sigma}_\oplus^n - \log\hat{\Sigma}_\oplus^a\right\rVert_F^2 + \left\lVert \log\hat{\Sigma}_\oplus^m - \log\hat{\Sigma}_\oplus^a\right\rVert_F^2.$$
Using 1000 bootstrap samples, the global $p$-value was found to be $p=0.015$. An alternative test, obtained by replacing the estimated Wasserstein covariance matrices with the corresponding correlations in the above statistic, resulted in $p = 0.093.$ 
\begin{figure}[t]
\centering
\includegraphics[scale = 0.7]{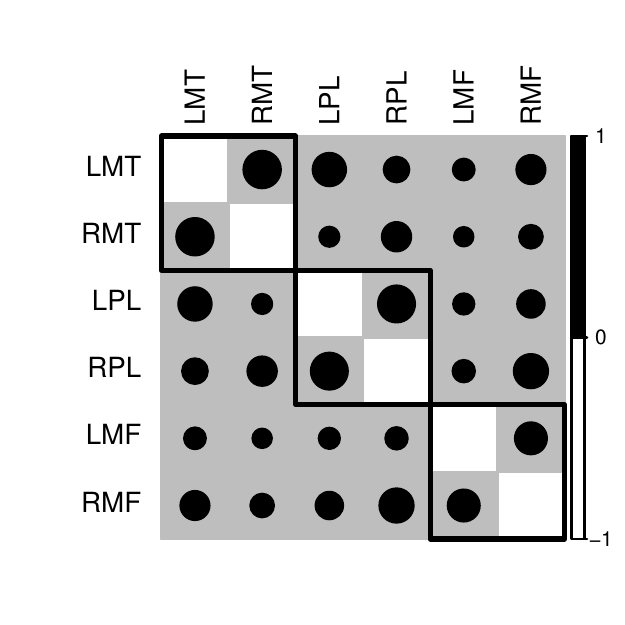} \quad
\includegraphics[scale = 0.7]{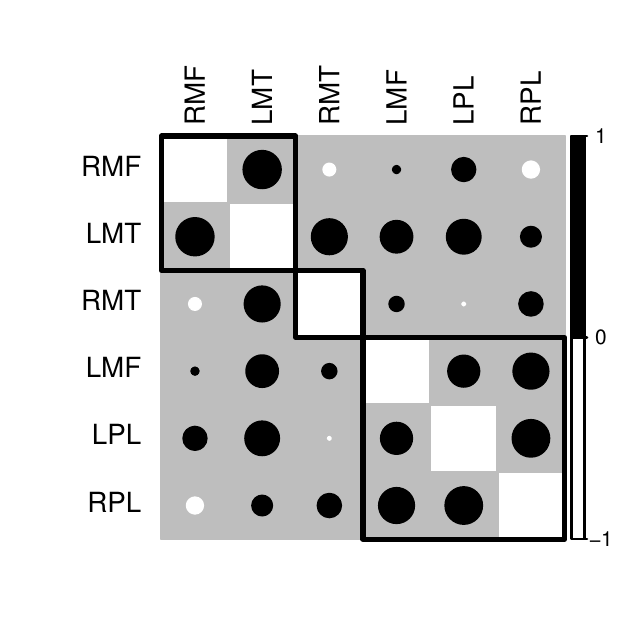} 
\caption{Estimated Wasserstein correlation submatrices corresponding to lateral hub pairs for normal (left) and Alzheimer's (right) subjects, after reordering of hubs by hierarchical clustering using Ward's criterion.  Rectangles indicate the groupings when three clusters are used.  Labels correspond to those in Figure~\ref{fig: wass_corrs} and further explanations can be found in the caption of Figure~\ref{fig: wass_corrs}.}
\label{fig: brain_clust}
\end{figure}
To further explore 
the differences between normal and Alzheimer's groups, the matrices were reduced to a subset of rows/columns corresponding to the lateral hub pairs in the middle frontal, middle temporal and parietal regions, then reordered using Ward's hierarchical clustering algorithm with three clusters, visualized in 
Figure~\ref{fig: brain_clust}. This  demonstrates the presence of  asymmetry 
in the Alzheimer's disease group, which is absent in the normal group.  \ci{derf:11} and others report findings of similar asymmetries in the brains of Alzheimer's patients. 

One can ask how the Wasserstein covariance approach compares to established approaches in brain imaging. A commonly used measure of local connectivity is regional homogeneity \cp{zang:04}, corresponding to Kendall's coefficient of concordance between the BOLD signals  at the seed voxel and those of its immediate neighbors.  This  scalar measure of regional homogenity can be computed for each hub and each subject, resulting in a sample of ordinary multivariate data.  The group covariance and correlation matrices are shown in the Supplementary Material.  Some common patterns are seen, most notably an increased number of negative correlations for the Alzheimer's group.  However, the differences are not as stark, and the regional homogeneity covariances do not reveal the asymmetry seen in the Wasserstein covariance analysis.

\subsection{Distribution of Age at Death for Period Cohorts}
\label{ss: mortality}

To gain a better understanding of  human longevity, the study of the temporal evolution of the distributions of age at death and their dependency structure over calendar time is of interest. The  Human Mortality Database provides 
yearly mortality and population data for 38 countries at \url{<www.mortality.org>}, which 
 have been previously analyzed with various functional data analysis techniques 
 \cp{hynd:09, mull:09:3}.

For a given country and calendar year, the probability distribution for mortality can be represented by its density. 
Consider a country for which life tables are available for the years $y_j  \, (j = 1, \ldots, p)$.  For integer-valued ages $a = 0, \ldots, 110$, the life table provides the size of the population $m_a$ at least $a$ years old, normalized so that $m_0 = 100000$.
These life tables were converted to histograms of age-at-death, which we then smoothed, applying the 
\texttt{hades} package, available at \url{<http://www.stat.ucdavis.edu/hades/>}, with a smoothing bandwidth of $h = 2$. This led to  estimated densities of age-at-death on the age domain  
$[20 \text{years}, 110 \text{years}]$. 

\begin{figure}[t]
\centering
\includegraphics[scale = 0.3]{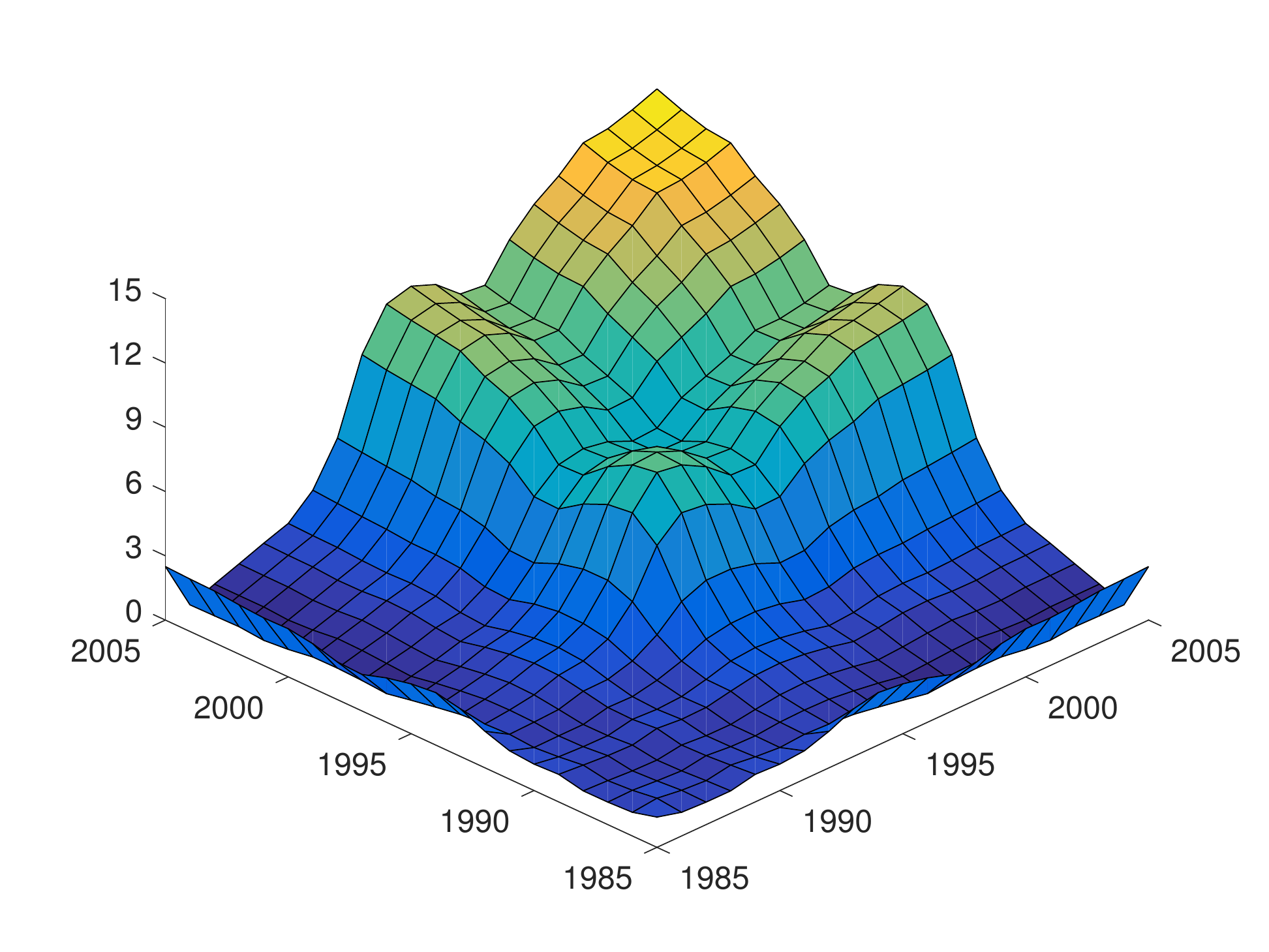} \quad
\includegraphics[scale = 0.3]{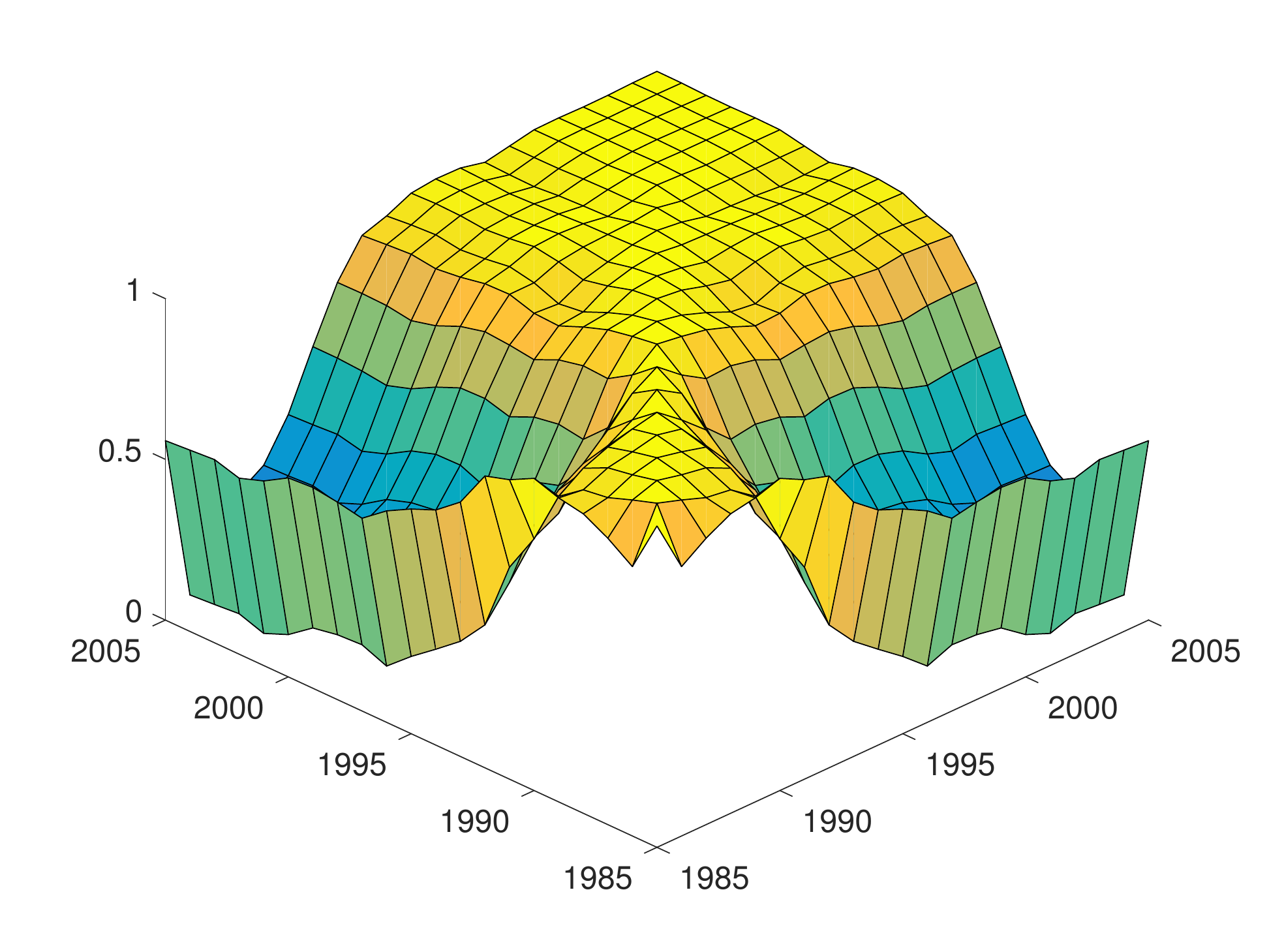} \\
\includegraphics[scale = 0.3]{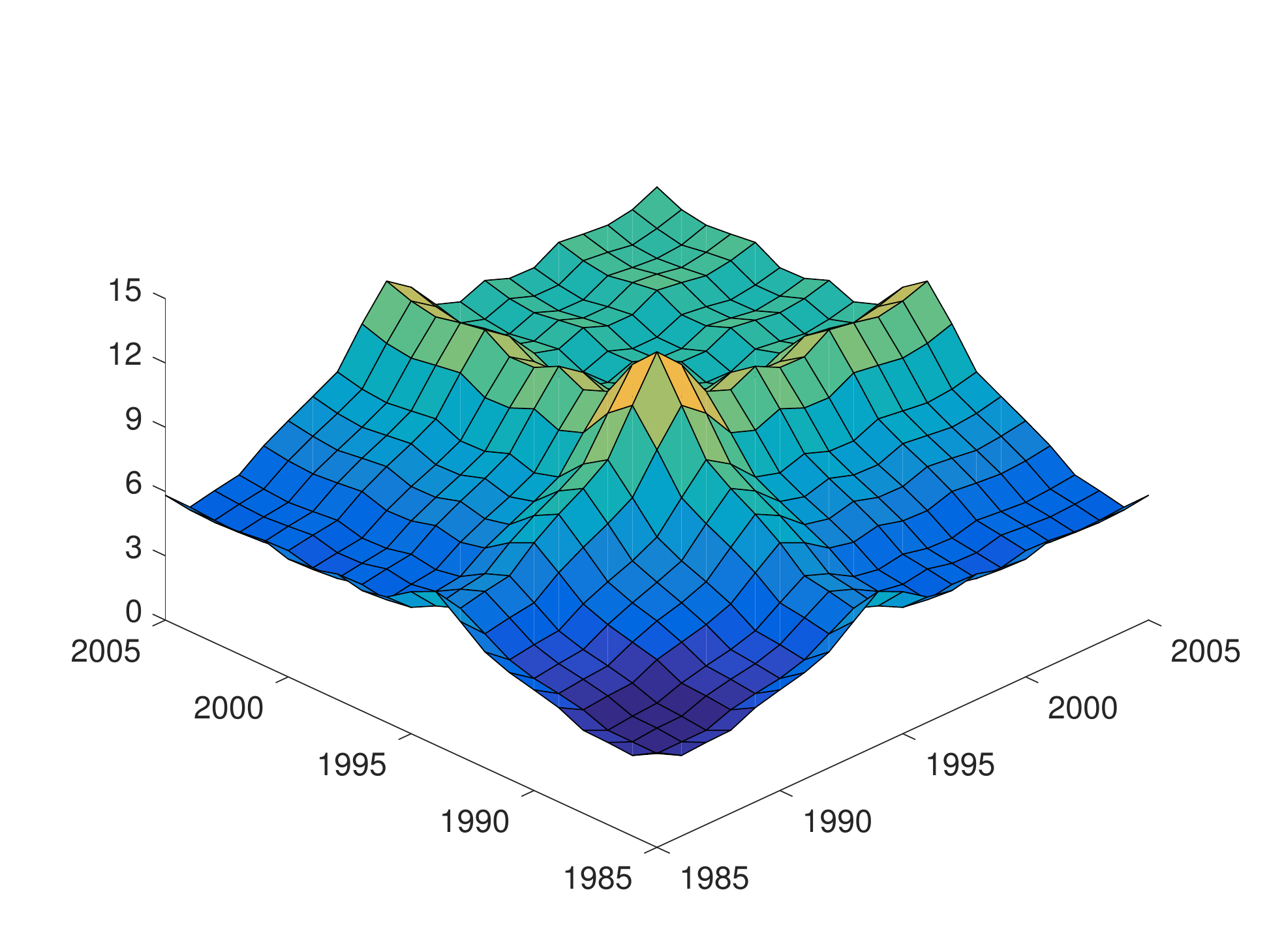} \quad
\includegraphics[scale = 0.3]{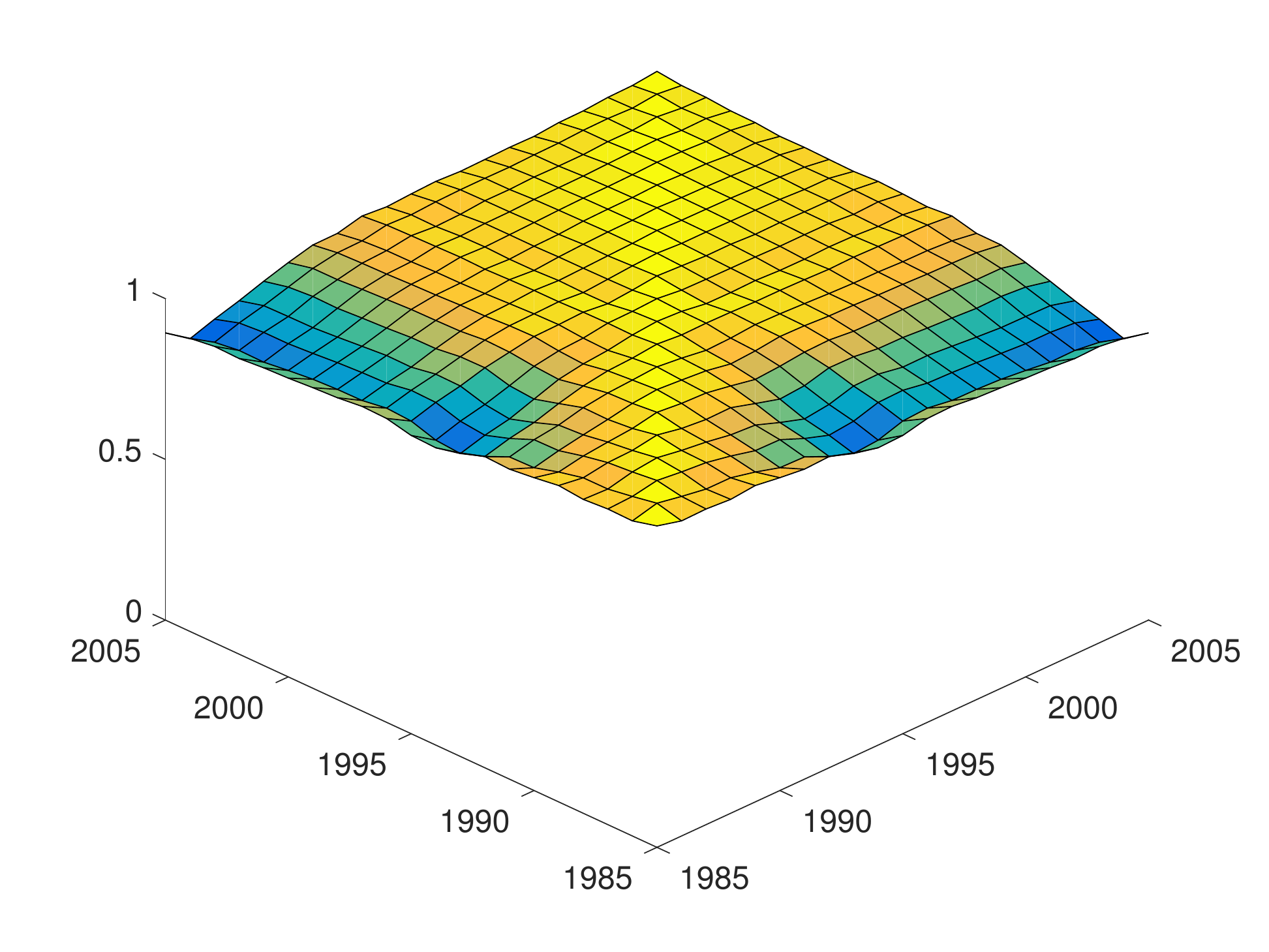}
\caption{Estimated Wasserstein covariance (left column) and correlation (right column) matrices for Eastern European (top row) and other (bottom row) countries. \label{fig: ww_ee_cov}}
\end{figure}

To illustrate the proposed methods for continuously varying densities, we considered  32 countries and identify  a subgroup of $n_E = 8$ countries located in Eastern Europe for comparison with the remaining $n_O = 24$ countries.  Densities were estimated for every year between 1985 and 2005. Wasserstein means are depicted in the Supplementary Material and are found to be quite similar for the two groups, demonstrating  increasing longevity over calendar time.  In contrast, the estimated Wasserstein covariance and correlation surfaces of these two groups differ quite drastically, as seen in Figure~\ref{fig: ww_ee_cov}.

Examining the diagonals of the Wasserstein covariance plots, the Eastern European countries are characterized by stagnant Wasserstein variability until 1993 when it increases sharply, followed by a steady increase 
between 1998 and 2005.  For the non-Eastern European countries, the Wasserstein variability is maximal in 1994. 
Wasserstein correlations reveal that mortality dependencies are  high and roughly constant over time for the non-Eastern European  countries, while they are weak for Eastern European countries between years before 1990 and those after 1995. 

These  exploratory findings are  not altogether surprising,  given the economic and societal upheaval in  Eastern Europe beginning around 1990, which is seen to be reflected in the Wasserstein covariance. 
In addition to exploratory analysis, we implemented a bootstrap test as described in the previous section.  With estimated Wasserstein covariance surfaces $\hat{\Sigma}_\oplus^E$ and $\hat{\Sigma}_\oplus^O$,  the test statistic was computed as the square root distance between the associated operators \cp{pigo:14} and implemented by computing the Frobenius distance between the principal square roots of the discretized matrix estimates.  Computing this statistic for 1000 bootstrap samples, the $p$-value for the difference between Wasserstein covariances of Eastern European and other countries was $0.287$, while it was $0.007$ using Wasserstein correlations,  providing some  evidence for differences in the distributions of the density processes.

\vspace{-0.6cm}
\section{Discussion}
\label{sec: discussion}

For studying the covariance structure of vectors of random densities, the Wasserstein approach is preferred over possible alternatives such as the transformation approach \cp{mull:16:1} or compositional methods based on the Aitchison geometry \cp{egoz:06}. This is because of  its convincing practical behavior for the construction of  barycenters \cp{bols:03,mull:11:4} and the theoretically appealing connections with optimal transport. 

Specifically, the transformation approach where densities are mapped to the entire Hilbert space $L^2$ by means of a suitable transformation such as the log-quantile density transformation, could be applied to the components of the vectors of densities, which would lead to unconstrained multivariate functional data. For such data, any one of numerous available measures of functional covariance and correlation  \cp{leur:93,mull:05:2,euba:08,mull:11:2} could then be harnessed. One would need to choose among many possible covariance measures and transformation maps, none of which is isometric to the Wasserstein distance. The resulting metric distortions make such an approach difficult to interpret.

In contrast, the proposed Wasserstein covariance has a canonical interpretation 
as an expected value of inner products of optimal transport maps.  This means  that the proposed Wasserstein covariance is similar to the notion of a regular covariance for an appropriate inner product, and can be interpreted as a measure of the degree of synchronization of the movement of probability mass from the Fr\'echet means to the random components of a bivariate density process. It thus emerges as a natural and compelling extension of the Wasserstein--Fr\'echet variance.
The quantile function representation of the Wasserstein covariance  in \eqref{eq: wass_quant} facilitates  the joint Wasserstein analysis of $p$ one-dimensional distributions  with Wasserstein covariance matrices and surfaces, enhancing the appeal of the proposed approach for practical applications.

While the geometric notion of covariance in \eqref{eq: wass_cov}
can be extended to the case where the sample densities have a  multivariate domain,  the  implementation via quantile functions cannot be extended for this case, and therefore the asymptotic theory we provide remains limited to the case of one-dimensional densities. An extension to multivariate domains and 
analogous notions of covariance with respect to other metrics are open problems for future research.

\end{document}